\documentclass[conference]{IEEEtran}
\IEEEoverridecommandlockouts
\usepackage{cite}
\usepackage{amsmath,amssymb,amsfonts}
\usepackage{algorithmic}
\usepackage{graphicx}
\usepackage{textcomp}
\usepackage{booktabs}

\usepackage{xcolor}
\def\BibTeX{{\rm B\kern-.05em{\sc i\kern-.025em b}\kern-.08em
    T\kern-.1667em\lower.7ex\hbox{E}\kern-.125emX}}
\begin{document}

\title{DiffChip: Thermally Aware Chip Placement with Automatic Differentiation\\

\thanks{This project was funded by the MIT-IBM Watson AI Lab.}
}

\author{\IEEEauthorblockN{Giuseppe Romano}
\IEEEauthorblockA{\textit{Institute for Soldier Nanotechnologies} \\
\textit{Massachusetts Institute of Technology}\\
Cambridge, USA \\
romanog@mit.edu}
\and
\IEEEauthorblockN{Aakrati Jain}
\IEEEauthorblockA{\textit{IBM T.J. Watson Research Center} \\
Yorktown Heights, NY, USA \\
Aakrati.Jain@ibm.com}
\and
\IEEEauthorblockN{Nima Dehmamy}
\IEEEauthorblockA{\textit{MIT-IBM Watson AI Lab} \\
Cambridge, MA, USA \\
Nima.Dehmamy@ibm.com}
\and

\IEEEauthorblockN{Cheng Chi}
\IEEEauthorblockA{\textit{IBM T.J. Watson Research Center} \\
Yorktown Heights, NY, USA \\
cchi@us.ibm.com}
\and
\IEEEauthorblockN{Xin Zhang}
\IEEEauthorblockA{\textit{IBM T.J. Watson Research Center} \\
Yorktown Heights, NY, USA \\
xzhang@us.ibm.com}
}

\maketitle

\begin{abstract}
Chiplets are modular integrated circuits that can be combined to form a larger system, offering flexibility and performance enhancements. However, their dense packing often leads to significant thermal management challenges, requiring careful floorplanning to ensure efficient heat distribution. To address thermal considerations, layout optimization algorithms concurrently minimize the total wirelength and the maximum temperature. However, these efforts employ gradient-free approaches, such as simulated annealing, which suffer from poor scaling and slow convergence. In this paper, we propose \texttt{DiffChip}, a chiplet placement algorithm based on automatic differentiation (AD). The proposed framework relies on a differentiable thermal solver that computes the sensitivity of the temperature map with respect to the positions of the chiplets. Regularization strategies for peak temperature, heat sources, and material properties enable end-to-end differentiability, allowing for gradient-based optimization. We apply \texttt{DiffChip} to optimize a layout where the total wirelength is minimized while keeping the maximum temperature below a desired threshold.  By leveraging AD and physics-aware optimization, our approach accelerates the design process of microelectronic systems, exceeding traditional trial-and-error and gradient-free methods.
\end{abstract}

\begin{IEEEkeywords}
2.5-D integrated circuits (ICs), chiplet floorplanning, computer architecture, hardware/software co-design.
\end{IEEEkeywords}

\section{Introduction}
The exponential increase in device density, as predicted by Moore's Law~\cite{moore1998cramming}, has been a key driver of technological advancement. However, this increase necessitates finer lithography techniques, leading to higher manufacturing costs. To address these limitations, chiplets have emerged as a promising solution. They enable efficient scaling by integrating heterogeneous components of varying complexities into a single package, enhancing flexibility while reducing manufacturing costs and complexity~\cite{douglas2017advanced,iyer2016heterogeneous}. Despite these advantages, chiplets present several challenges such as interconnect complexity~\cite{ho2013multiple} and thermal dissipation~\cite{skadron2003temperature,torregiani2009thermal}. As these issues are greatly influenced by the chiplet locations, floorplanning optimization is a crucial design step. Chiplet placement may take into account several factors, such as maximum wirelength and data delay~\cite{chen2023floorplet,zhuang2022multi}, as well as thermal effects~\cite{han20222,wang2023efficient}. Typically, thermally-aware chiplet floorplanning uses \emph{gradient-free} optimization, where the physics of the problem is treated as a black box. A popular heat conduction solver employed in this class of methods is HotSpot~\cite{zhang2015hotspot}. Although these approaches provide insights into the response of the temperature profile to design choices, they become expensive for non-regular realistic layouts~\cite{eris2018leveraging}. To mitigate this issue, machine-learning techniques have been proposed. For instance, Ref.~\cite{chen2022fast} uses Graph Neural Networks (GNN) as a surrogate for the thermal solver to predict temperature profiles faster than HotSpot. However, the training cost undermines computational efficiency~\cite{woldseth2022use}.

This paper introduces \texttt{DiffChip}, a framework for chiplet layout optimization based on automatic differentiation (AD), written in JAX~\cite{jax2018github}. AD enables faster development iterations by taking gradients of programs without manual sensitivity analysis. It is fundamental in machine learning for calculating gradients with respect to neural network weights, often using reverse-mode AD, or backpropagation. Beyond machine learning, AD is increasingly used in physics-based inverse design~\cite{romano2022inverse,hammond2021photonic}, where the cost function often depends on the solution of a partial differential equation (PDE). In such a case, the optimization is said \emph{PDE-constrained}. Our work falls under this category. The main challenge is to differentiate through the chiplet positions. In fact, from a modeling standpoint, chiplets are rectanguloid to which we assign a given thermal conductivity and volumetric power density. Thus, there is a spatial discontinuity of these quantities, which breaks the differentiability of the pipeline as the chiplets \emph{move} along the plane during optimization. We address this issue by adopting smoothing strategies. Our contributions are summarized below:

\begin{itemize}

   \item The development of a smoothing model for the thermal conductivity and heat source associated to chiplets. 
   \item The regularization of the maximum temperature (which is not differentiable) into a differentiable form enables the calculation of its sensitivity with respect to the chiplet locations. 
   \item The development of a novel differentiable non-overlap chiplet constraint. As explained in the Sec.\ref{nooverlap}, this constraint is based on smoothed \emph{indicator} functions.
   \item The implementation of a regularization for the total wirelength.
   \item The development of a code that integrates these abovementioned regularized quantities with a differentiable 3D thermal solver.
   \item An example of layout optimization, where the HPWL is minimized under maximum temperature and non-overlap constraints. 
   \item The validation of the optimized layout with a commercial software.
   
\end{itemize}

This paper is organized as follows. In Sec.~\ref{sec:statement}, the problem statement is outlined, followed by the description of regularization strategies (\ref{regularization}) pertaining to the thermal conductivity, heat source, the HPWL and the maximum temperature. In Sec.~\ref{nooverlap} the non-overlap constraint is detailed. Section~\ref{results} shows the application of \texttt{DiffChip} on two optimization scenarios, followed by conclusions (Sec.~\ref{conclusions}).

\section{Problem statement}\label{sec:statement}

\begin{figure}[h!]
    \centering
    \includegraphics[width=0.6\linewidth]{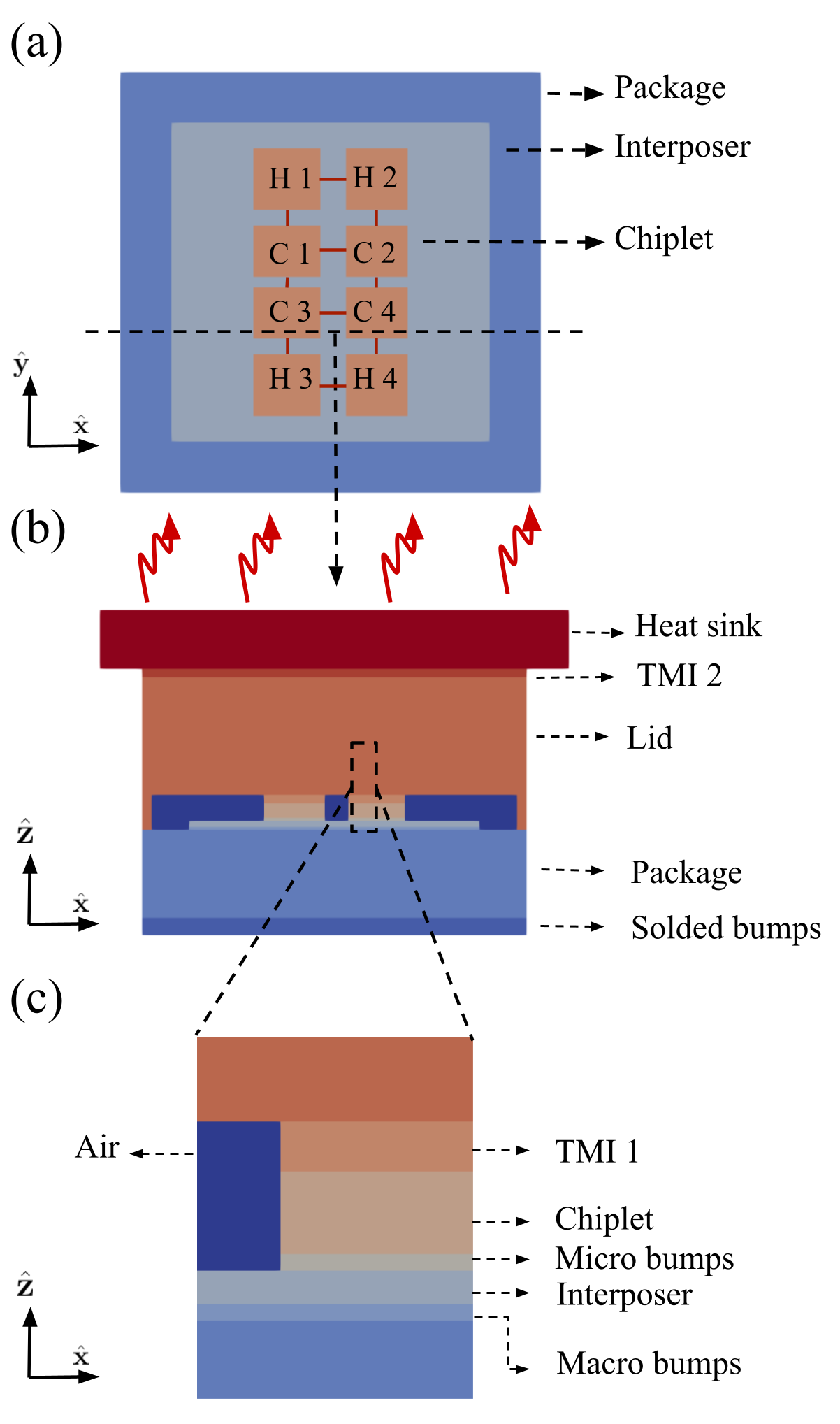}
    \caption{(a). Top view on the chiplet configurations. The connections among chiplets are shown in red. The lid and heat sink are omitted. (b) Vertical cut where the z-dimension is magnified 10x for clarity. (c) Close-up of the chiplet portion of the simulation domain.}
    \label{fig:fig1}
\end{figure}

\begin{figure}[h!]
     \centering
     \includegraphics[width=1\linewidth]{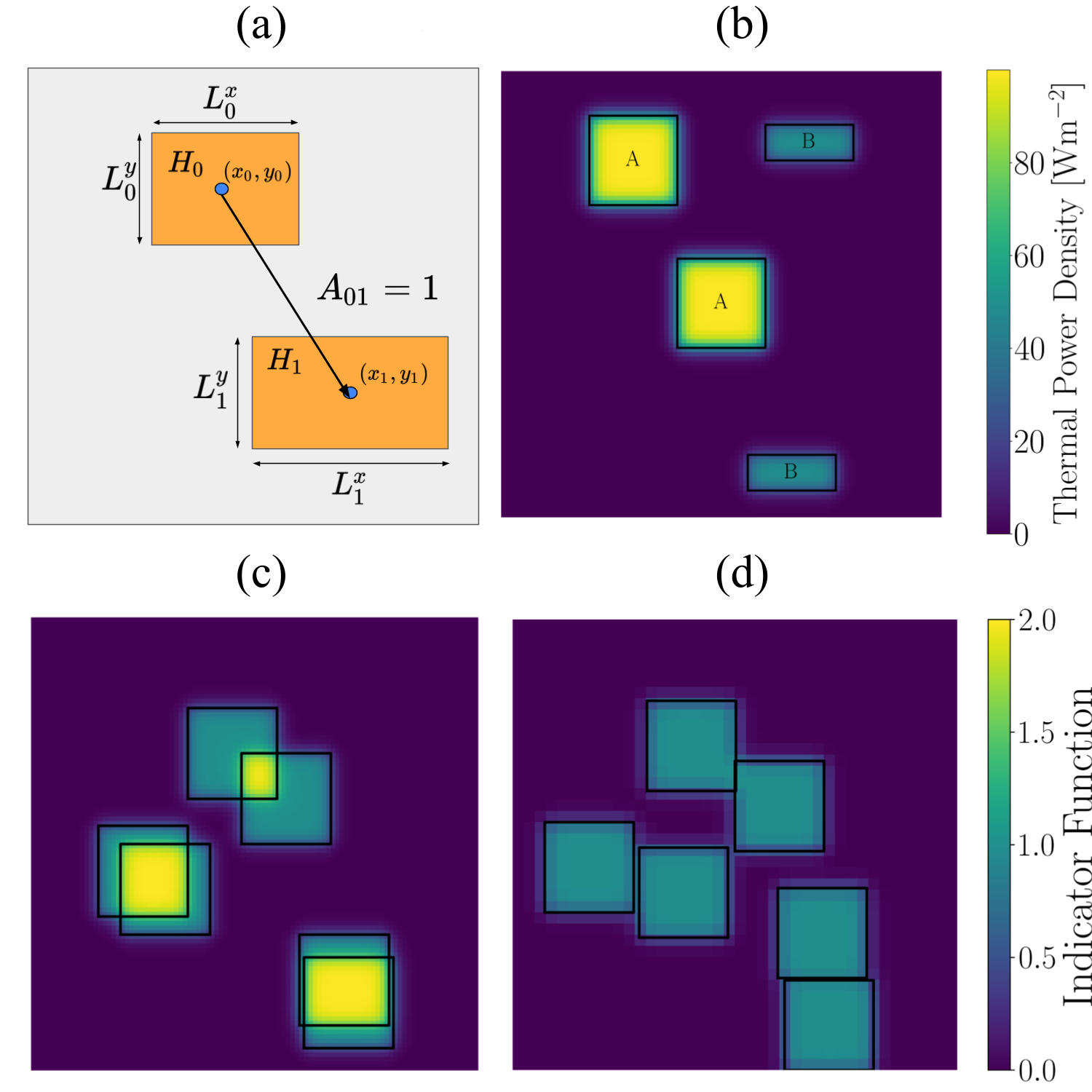}
     \caption{(a) Sketch depicting two connected chiplets and their parameters. (b) Example pf spatial power density distribution associated with a set of chiplets. (c) The initial chiplets configuration for the optimization problem from Eq.~\ref{opti_nooverlap}. (d) The optimized chiplets positions.}
     \label{fig:fig2}
\end{figure}

We consider a standard package architecture~\cite{jain2023thermal}, in which chiplets are bonded to the interposer via microbumps, and macrobumps connect the interposer to the package (Fig.~\ref{fig:fig1}). The package is secured with soldered bumps. To efficiently dissipate heat, a copper lid and heat sink are included. Heat dissipation is managed using a convective boundary condition applied to the top surface of the heat sink. Thermal interface materials are included between the chiplets and the lid, as well as between the lid and the heat sink. For complex geometries, such as the bumps and interposer, a homogenized anisotropic thermal conductivity is used. The sizes of each layer, along with their thermal conductivity tensors, are reported in Table~\ref{tab:values}. The goal is to minimize the total wirelength while maintaining the maximum temperature below a specified threshold, $T_{th}$, and ensuring chiplets do not overlap. To this end, we consider $N_c$ = 8 chiplets (shown in Fig.~\ref{fig:fig1}a), including four high-bandwidth memory (H) blocks and four compute (C) blocks. The connections among chiplets are described by the adjacency matrix $\mathbf{A}$, which is 1 only when chiplet $i$ is connected with chiplet $j$ and zero otherwise (Fig.~\ref{fig:fig2}a). Note that connections are intended as unidirectional; thus $A_{ij} = 1$ does not imply that $A_{ji} = 1$. In our example, as illustrated in Fig.~\ref{fig:fig1}a, there are 10 connections. Chiplets are assigned a dissipated power, $H_i$, and dimensions $L_i^x,L_i^y$, where $i = 0,...,N_c-1$ (Fig.~\ref{fig:fig2}b); the thickness of each die is fixed (0.25 mm). The positions of the chiplets are denoted by $\mathbf{p}\in\mathcal{R}^{N_c\times 2}$, comprising $2N_c$ degrees of freedom. The total wirelength is estimated by the Half Perimeter Wirelength (HPWL)~\cite{ray2014half}, formulated as~\cite{ray2014half} 
\begin{align}\label{hpwl}
\textrm{HPWL} &= \sum_{i,j=0}^{N_c - 1} A_{i,j} 
\left[ \max(x_i, x_j) - \min(x_i, x_j) \right. \nonumber \\
&\quad \left. + \max(y_i, y_j) - \min(y_i, y_j) \right],
\end{align}
while the temperature map is evaluted by the standarnd heat conduction equation
\begin{align}\label{fourier}
-\nabla \cdot \kappa \nabla T &= H && \mathbf{r} \in \Omega \nonumber \\
-\kappa \nabla T &= h \left(T - T_{\mathrm{amb}} \right) && \mathbf{r} \in \Gamma_1 \nonumber \\
-\kappa \nabla T &= 0 && \mathbf{r} \in \Gamma_2.
\end{align}
In Eq.~\ref{fourier}, the convective heat transfer coefficient is $h = 3\times 10^3$ Wm$^{-2}$ K$^{-1}$, the ambient temperature $T_{\mathrm{amb}}=20^\circ$C, $\Omega$ is the simulation domain, $\Gamma_1$ is the top side of the sink and  $\Gamma_2$ is the remaining of the boundary, which is treated as adiabatic. Equation~\ref{fourier} is discretized using the finite-volume method, resulting in the following linear system
\begin{equation}\label{discretized}
\mathbf{A} \mathbf{T} =\mathbf{b},
\end{equation}
where $\mathbf{A}\in \mathcal{R}^{N\times N}$, $\mathbf{T}\in \mathcal{R}^{N}$, $\mathbf{b}\in \mathcal{R}^{N}$, and $N = N_x N_y N_z$ is the number of control volumes. In this work, the grid is $N_x=100, N_y = 100$ and $N_z = 222$, amounting to 2.22 million degrees of freedom. The higher resolution along the $z$-axis accounts for the small thickness of layers compared to their in-plane dimensions. The control volume has the dimension $0.8\times0.8\times0.025$ mm$^3$. 

\begin{table}[h!]
\renewcommand{\arraystretch}{1.3} 
\centering
\begin{tabular}{@{}lcc@{}}
\toprule
\textbf{Layer} & \multicolumn{1}{c}{\textbf{Thermal Conductivity}} & \multicolumn{1}{c}{\textbf{Size}} \\ 
               & \multicolumn{1}{c}{($\kappa_{xx}$, $\kappa_{yy}$, $\kappa_{zz}$)} & \multicolumn{1}{c}{($L_x$, $L_y$, $L_z$)} \\ 
               & \multicolumn{1}{c}{[Wm$^{-1}$K$^{-1}$]} & \multicolumn{1}{c}{[mm]} \\ \midrule
Heat Sink      & (385,385,385)          & (80,80,1)             \\ \midrule                
TMI 2          & (1.6,1.6,1.6)          & (66,66,0.15)          \\ \midrule                 
Lid            & (385,385,385)          & (66,66,2.0)             \\ \midrule                
TMI 1          & (1.6,1.6,1.6)          & ($L_x$,$L_y$,0.15)    \\ \midrule                 
Chiplet        & (150,150,150)          & ($L_x$,$L_y$,0.25)    \\ \midrule                 
Micro bumps    & (0.9,0.9,2.5)          & ($L_x$,$L_y$,0.05)    \\ \midrule                        
Interposer     & (128,128,148)          & (50,50,0.1)           \\ \midrule               
Macro bumps    & (0.9,0.9,2.5)          & (50,50,0.05)          \\ \midrule
Substrate      & (0.33,0.33,0.33)       & (66,66,1.5)           \\ \midrule
Solded bumps   & (1.2,1.2,2.8)          & (66,66,0.3)           \\ \midrule
Air     & (0.024,0.024,0.024)    & ----           \\ \bottomrule
\end{tabular}
\vspace{10pt} 
\caption{Size and thermal conductivity tensors of each layer, and for air. The thickness of the lid is 1 mm along all three directions. }
\label{tab:values}
\end{table}

\begin{table}[h!]
\renewcommand{\arraystretch}{1.3} 
\centering
\begin{tabular}{@{}lccc@{}} 
\toprule
\textbf{Chiplet} & \multicolumn{1}{c}{\textbf{Position}} & \multicolumn{1}{c}{\textbf{Size}} & \multicolumn{1}{c}{\textbf{Power}} \\ 
                 & \multicolumn{1}{c}{($x$, $y$)}       & \multicolumn{1}{c}{($L_x$, $L_y$)}& \multicolumn{1}{c}{$H$} \\ 
                 & \multicolumn{1}{c}{[mm]}             & \multicolumn{1}{c}{[mm]}          &    \multicolumn{1}{c}{[W]}                       \\ \midrule
H 1           &        (-7,16)       &  (10,10)              &  20                          \\ \midrule                
H 2           &        (7,16)       &   (10,10)            &     20                     \\ \midrule                 
H 3           &        (-7,-16)     &     (10,10)          &         20                 \\ \midrule                
H 4           &        (7,-16)     &     (10,10)          &          20                \\ \midrule                 
C 1           &        (-7,5)       &    (10,8)           &           30               \\ \midrule                 
C 2           &        (7,5)              &    (10,8)           &        30                  \\ \midrule                        
C 3           &        (-7,-5)               &   (10,8)            &        30                  \\ \midrule               
C 4           &       (7,-5)               &      (10,8)         &            30              \\ \bottomrule
\end{tabular}
\vspace{10pt} 
\caption{Chiplet layout and properties.}
\label{tab:chiplets}
\end{table}

\section{Regularization}\label{regularization}
Equation~\ref{hpwl} needs to be regularized because the maximum and minimum functions are not differentiable. To this end, we use the logsumexp softmax/min functions
~\cite{ray2014half}
\begin{eqnarray}
\max(\mathbf{x}) &=& \gamma \log \left[\sum_i e^{x_i \gamma}\right]  \\
\min(\mathbf{x}) &=& -\gamma \log \left[\sum_i e^{-x_i \gamma}\right],
\end{eqnarray}
with $\gamma = 10^{-3}$. Using these expressions, the HPWL becomes
\begin{eqnarray}\label{hpwl2}
\textrm{HPWL} = \gamma \sum_{i,j=0}^{N_c - 1} A_{i,j} 
\log\left[2 + 2 \cosh\left(\frac{x_j - x_i}{\gamma}\right)\right] + \nonumber \\ + A_{i,j} 
\log\left[2 + 2 \cosh\left(\frac{y_j - y_i}{\gamma}\right)\right].
\end{eqnarray}
To regularize the maximum temperature from Eq.~\ref{discretized}, we use the p-norm approximation, which is commonly used in thermal inverse problems~\cite{lohan2020study},
\begin{equation}\label{pnorm}
\max \mathbf{T} \approx \left(\frac{1}{N}\sum_{i=0}^{N-1} T_i^{p}\right)^{1/p} = \text{p-norm}(\mathbf{T}),
\end{equation}
with $p$ = 90. Another term that needs regularization is the heat source $H$ in Eq.~\ref{fourier}. In fact, assuming spatially uniform power dissipation, each chiplet is assigned a 3D box function, which is not differentiable. 

\begin{figure*}[h!]
    \centering
    \includegraphics[width=0.75\textwidth]{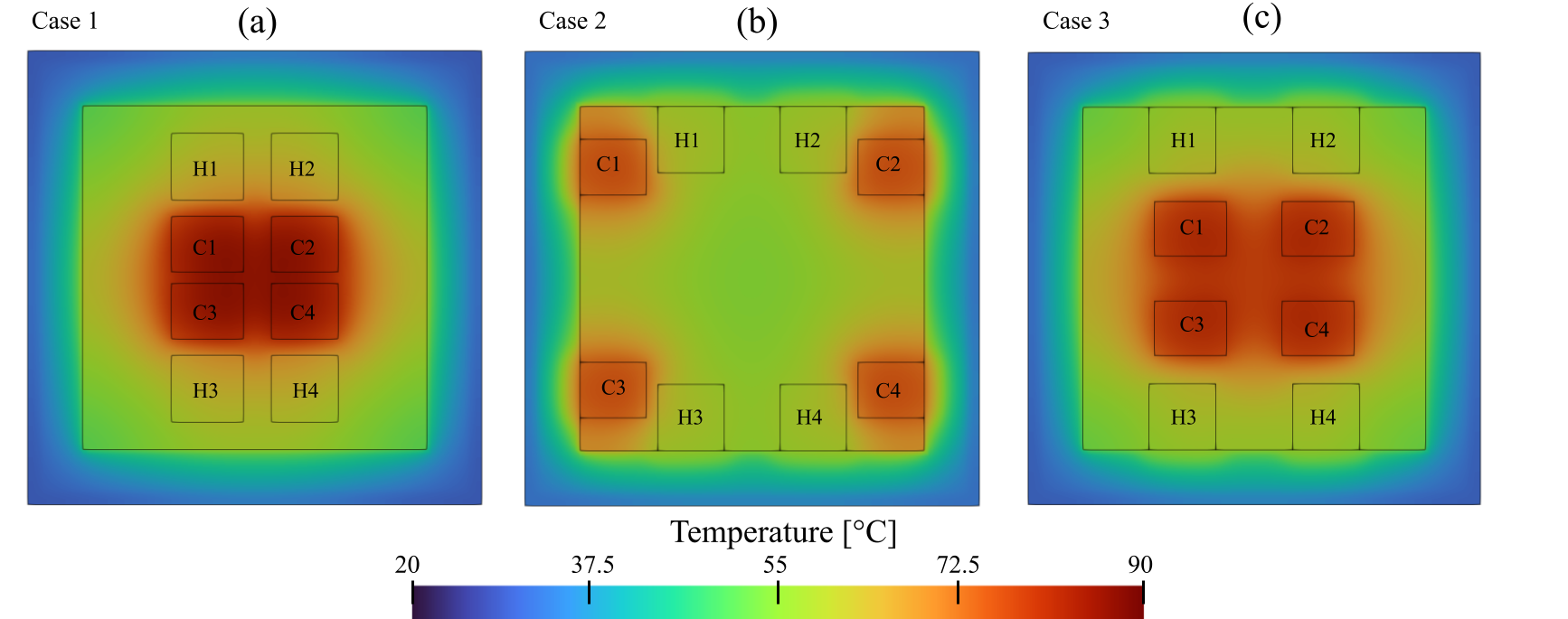}
    \caption{(a) Temperature map for the initial configuration (Case 1), comprising four high-bandwidth and four compute units. The peak temperature is 85.98°C. (b) Temperature distribution for the case where the maximum temperature is minimized without HPWL constraint (Case 2), with a peak of 75.44°C. (c) The configuration that minimizes the HPWL while keeping the temperaturte below 82°C (Case 3). The peak temperature is 80.81°C.}
    \label{fig:map}
\end{figure*}
The key idea is to smooth out this box function along the \emph{xy} plane, while keeping the discontinuity along $z$, where differentiability is not needed. The first step is to express the heat source as
\begin{eqnarray}
H(x,y,z) = \mathcal{H}^z(2|z-z_c|/L^z) \times \nonumber \\ \sum_i H_i\mathcal{H}^{\parallel}(2|x-x_i|/L_x^i,2|y-y_i|/L_y^i) ,
\end{eqnarray}
where $\mathcal{H}^{\parallel}$ and $\mathcal{H}^z$, are the in-plane 2D and out-of-plane 2D Heaviside functions. The term $z_c$ is the midpoint in the $z$ direction of the chiplets. Next, we regularize the in-plane Heaviside functions with 
\begin{eqnarray}\label{indicator}
\Theta_i(x,y) = \frac{1}{4}\left[\tanh\left({\alpha|x_i - x| - \alpha L^x_i/2}\right) + 1\right] \nonumber \times \\ \times \left[\tanh\left(\alpha{|y_i - y| - \alpha L^y_i/2}\right) + 1\right],
\end{eqnarray}
where $\alpha = 1$ is a smoothing parameter. An example of power dissipation map is shown in Fig.~\ref{fig:fig1}-b. A similar regularization approach is applied to the space-dependent thermal conductivity. Specifically, each \emph{movable} block is composed of a microbump layer, a die and a thermal material interface. Their corresponding thermal conductivity distributions need to be differentiable along the $xy$-plane. Let's refer to the thickness, the midpoint in their $z$ direction, and thermal conductivity of each of these three layers as $t_l$, $z_l$ and $\kappa_l$, respectively, with $l=0,1,2$. Then, the space-dependent thermal conductivity is
\begin{eqnarray}\label{kappa}
\kappa(x,y,z) = \kappa_B(x,y,z) +  \sum_{i=0}^{N_c-1} \Theta_i(x,y) \times \nonumber \\ \times\sum_{l=0}^2(\kappa_l-\kappa_{\mathrm{air}})\mathcal{H}^z(2|z-z_l|/t_l),
\end{eqnarray}
where $\kappa_B$ is the thermal conductivity of the fixed part of the simulation domain; as this term also includes the air enclosed by the lid, in Eq.~\ref{kappa}, we subtract $\kappa_{\mathrm{air}}$ from $\kappa_l$.

\section{non-overlap constraint}\label{nooverlap}
During optimization, chiplets may overlap; to avoid such a scenario, we introduce an inequality constraint based on the following rationale: Suppose we have a 2D function that is 1 only at the chiplet locations and zero elsewhere, the maximum of such a function becomes larger than one when overlap occurs. To make such a constraint differentiable, we combine two concepts discussed in the last section. First, we build the smooth 2D function
\begin{equation}
f(x,y) = \sum_{i=0}^{N_c-1}\Theta_i(x,y),
\end{equation}
and then compute the maximum value as $\textrm{p-norm}(f)$. The non-inequality constraint is
\begin{equation}\label{inequality2}
g_0(\mathbf{p}) = \left[ \frac{1}{N}\sum_{k=0}^{N -1}f_k^{p}\right]^{1/p} -1\le 0.
\end{equation}
with $p = 90$. Note that we used the discretized version of $f$, given by $f_k = f(x_k,y_k)$. To validate this approach, we solve the following optimization problem
\begin{equation}\label{opti_nooverlap}
\min_{\mathbf{p}}g_0(\mathbf{p}).
\end{equation}
 Throughout this work, as an optimizer, we choose the Method of Moving Asymptotes (MMA)~\cite{svanberg1987method}. The starting guess, shown in Fig.~\ref{fig:fig2}-c, is a set of partially overlapped chips. As illustrated in Fig.~\ref{fig:fig2}-d after roughly 20 iterations, the chiplets do indeed spread zeroing the overlap.

\section{Results}\label{results}

\begin{figure}[h!]
    \centering
    \includegraphics[width=0.4\textwidth]{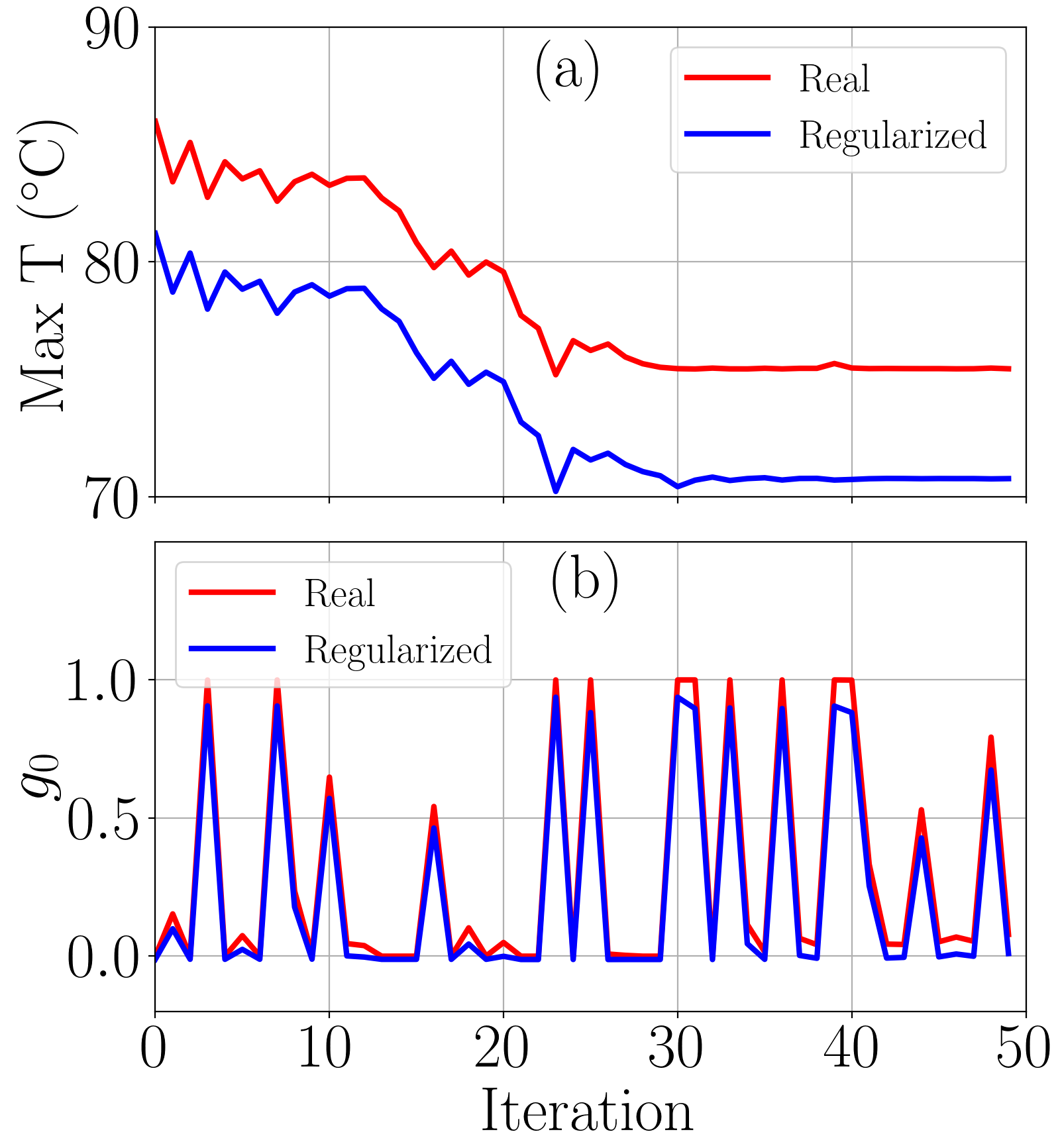}
    \caption{(a) The evolution of the p-norm and real maximum temperature for Case 2, where there is no constraint on the HPWL. The two quantities follow the same trend and are consistently about 5°C apart, corroborating the use of the pnorm as a proxy for the maximum temperature (b) The evolution of the non-overlap constraint $g_0$ from Eq.~\ref{inequality2}. It follows closely its real counterpart.}
    \label{fig:case_2}
\end{figure}

We first consider the configuration shown in Fig.~\ref{fig:fig1}a, where the regularized (from Eq.~\ref{hpwl}) and the real HPWL are 92 and 103.1 mm, respectively (Tab. \ref{tab:HPWL}). We refer to this scenario as ``Case 1.'' The temperature map, shown in Fig.~\ref{fig:map}a, is higher in the central area, where the high-power blocks (compute units) are located. The peak value is 85.98\textdegree C, while the regularized one, computed by Eq.~\ref{pnorm}, is 81.20\textdegree C. For comparison, we also compute the temperature map with \texttt{ANSYS}, which predicts a temperature peak of 86.85\textdegree C, falling within 1\% error.
Next, to test the optimization algorithm, we minimize the peak temperature while avoiding chiplet overlap (Case 2). The first guess is given by the configuration in Fig.~\ref{fig:fig1}a. The optimization algorithm is
\begin{eqnarray}\label{eq:algo_1}
&&\min_{\mathbf{p}} \text{p-norm}(\mathbf{T}) \\
&& \textrm{s.t.}\,\,\, \mathbf{T}  = \mathbf{A}^{-1}\mathbf{b} \\
&& \textrm{s.t.}\,\,\, g_0(\mathbf{p}) \le 0.
\end{eqnarray}
As illustrated in Fig.~\ref{fig:case_2}a, the real maximum temperature decreases by approximately 10°C over 30 iterations, reaching a peak temperature of 75.44°C. In the final arrangement depicted in Fig.~\ref{fig:map}b, the chiplets are positioned near the border, with compute units spaced farther apart. This layout is expected, as components with higher power tend to be more susceptible to thermal crosstalk. We note that the final layout maintains the symmetry of the initial guess since the convective boundary condition applied at the top surface is applied uniformly. The regularized temperature, also shown in Fig.~\ref{fig:case_2}a, remains consistently lower by about 5°C, closely mirroring the trend of the actual temperature. This observation corroborates the effectiveness of using the p-norm as a representation of the maximum temperature. For this configuration, \texttt{ANSYS} predicts a peak temperature of 76.33°C, which, again, is within 1\% error. 
\begin{table}[h!]
\renewcommand{\arraystretch}{1.1} 
\centering
\begin{tabular}{@{}lccc@{}} 
\toprule
\textbf{Solver} & \multicolumn{1}{c}{\textbf{Case 1}} & \multicolumn{1}{c}{\textbf{Case 2}} & \multicolumn{1}{c}{\textbf{Case 3}} \\ 
                
                  & \multicolumn{1}{c}{[C\textdegree]}       & \multicolumn{1}{c}{[C\textdegree]}   &  \multicolumn{1}{c}{[C\textdegree]}  \\ \midrule
\texttt{DiffChip} (p-norm)         &        81.20             &      70.77                   &   76.28                        \\ \midrule 
\texttt{DiffChip} (real)         &        \textbf{85.98}      &      \textbf{75.44}          &   \textbf{80.81}                        \\ \midrule                
\texttt{ANSYS}                   &         86.85             &   76.33            &      81.67                            \\ \bottomrule               

\end{tabular}
\vspace{10pt} 
\caption{Peak Temperatures for different cases.}
\label{tab:temperature}
\end{table}
\begin{table}[h!]
\renewcommand{\arraystretch}{1.1} 
\centering
\begin{tabular}{@{}cccc@{}} 
\toprule
\textbf{HPWL} &\multicolumn{1}{c}{\textbf{Case 1}} & \multicolumn{1}{c}{\textbf{Case 2}} & \multicolumn{1}{c}{\textbf{Case 3}} \\ 
                
& \multicolumn{1}{c}{[mm]}       & \multicolumn{1}{c}{[mm]} &       \multicolumn{1}{c}{[mm]}               \\ \midrule
\texttt{Softmax}    &    92      &      203.5          &    120.8  \\ \midrule

Real    &   \textbf{103.1}      &      \textbf{209.2}          &    \textbf{128.6}                              \\ \bottomrule               

\end{tabular}
\vspace{10pt} 
\caption{The half-permimeter wirelength (HPWL) for different cases.}
\label{tab:HPWL}
\end{table}
\begin{table}[h!]
\renewcommand{\arraystretch}{1.3} 
\centering
\begin{tabular}{@{}lccc@{}} 
\toprule
\textbf{Chiplet} & \multicolumn{1}{c}{\textbf{Case 1}} & \multicolumn{1}{c}{\textbf{Case 2}} & \multicolumn{1}{c}{\textbf{Case 3}} \\ 
                 & \multicolumn{1}{c}{($x$, $y$)}       & \multicolumn{1}{c}{($x$, $y$)}& \multicolumn{1}{c}{($x$, $y$)} \\ 
                 & \multicolumn{1}{c}{[mm]}             & \multicolumn{1}{c}{[mm]}          &    \multicolumn{1}{c}{[mm]}                       \\ \midrule
H 1           &        (-7,16)       &    (-8.93,20.)           &   (-10.21,20.)                         \\ \midrule                
H 2           &        (7,16)       &    (8.93,20.)           &        (10.21,20)                  \\ \midrule                 
H 3           &        (-7,-16)     &   (-8.93,-20)           &        (-10.21,-20)                    \\ \midrule                
H 4           &        (7,-16)     &      (8.93,-20)            &         (10.21,-20)                \\ \midrule                 
C 1           &        (-7,5)       &       (-19.9, 15.83)      &         (-9.07, 6.85)                \\ \midrule                 
C 2           &        (7,5)              &  (19.9, 15.83)        &      (9.07, 6.85)                    \\ \midrule                 
C 3           &        (-7,-5)               &     (-19.9 ,-15.83)            &   (-9.07,-6.85)                       \\ \midrule 
C 4           &       (7,-5)               &    (19.9 ,-15.83)         &    (9.07,-6.85)              \\ \bottomrule
\end{tabular}
\vspace{10pt} 
\caption{Chiplet layout for different cases.}
\label{tab:chiplets2}
\end{table}
\begin{figure}[h!]
    \centering
    \includegraphics[width=0.4\textwidth]{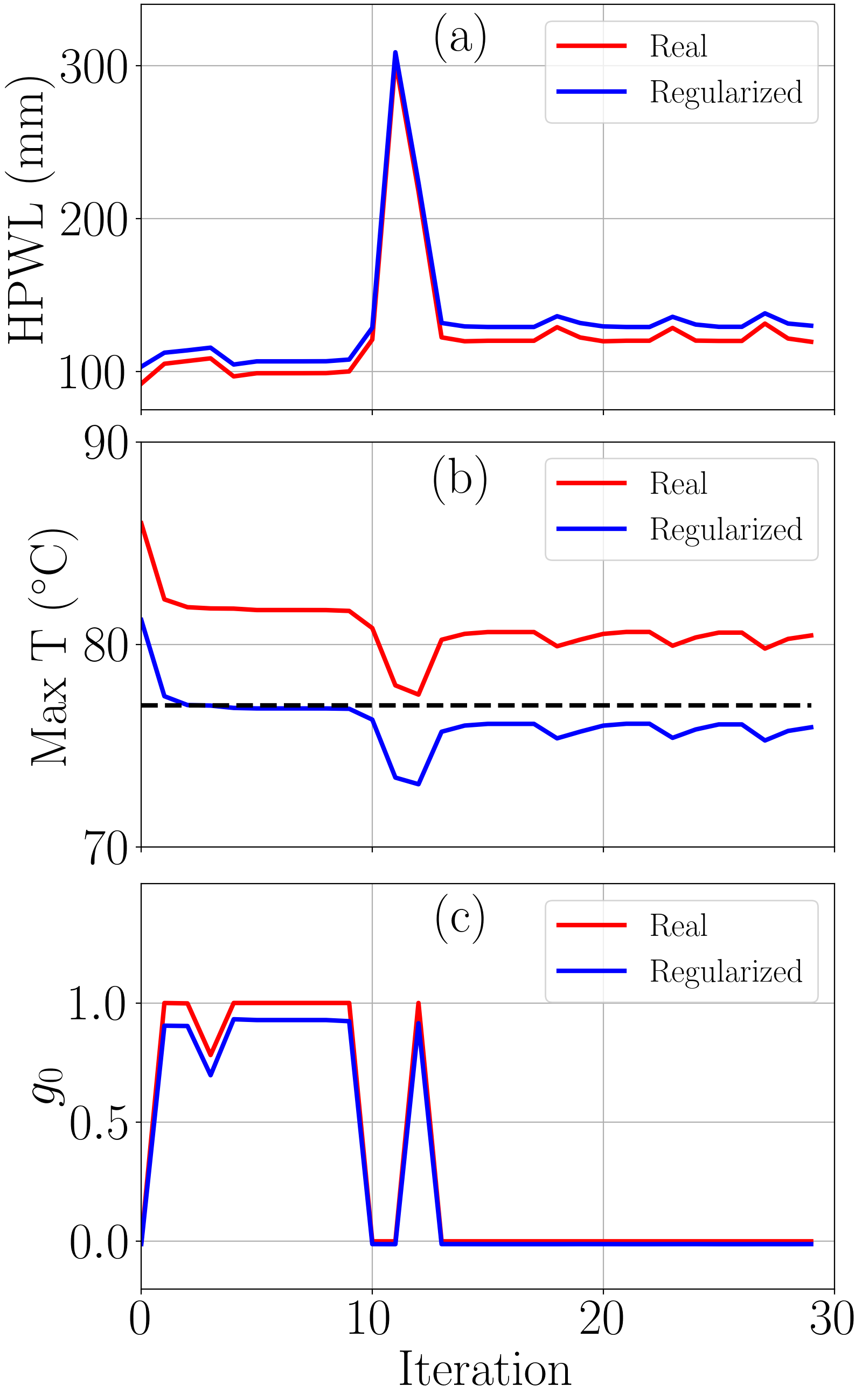}
    \caption{(a) The evolution of the HPWL (both the real and the regularized one). It increases as the optimization progresses in order to accomodate the constraint on the maximum temperature. (b) Trajectory of the maximum temperature. At convergence, the regularized temperature falls below $T_{th}$=75°C (c) The evolution of the nonoverlap constraint. After an initial phase, where it is violated, both the real and the regularized constraint approche zero. i.e. the constraint it satisfied.}
    \label{fig:case_3}
\end{figure}
The evolution of the nonoverlap constraint, as depicted in Fig.~\ref{fig:case_2}b, is oscillatory. This indicates that the optimizer can rapidly reposition chiplets to reduce overlap without substantially affecting the cost function. However, this is not the case when minimizing the HPWL, as will be discussed later. Since in this case there is no constraint on the HPWL, its final value is as large as 209.2 mm, with the regularized counterpart being 203.5 mm.  Finally, as illustrated in Fig.~\ref{fig:case_2}b, the regularized nonoverlap constraint closely matches its nondifferentiable counterparts, achieved by substituting the p-norm with the maximum value in Eq.~\ref{inequality2}.
\\
As the last example (Case 3), we consider the case where HPWL is minimized while constraining the p-norm temperature below a threshold, and ensuring no overlap. Supposed we want to maintain the peak temperature below $82$\textdegree C. From Case 2, we noted that there is a gap of about 5\textdegree C between the real and regularized peak temperature. Thus, we set $T_{th}$ = 77\textdegree C. The corresponding algorithm reads
\begin{eqnarray}\label{eq:algo_2}
&&\min_{\mathbf{p}} \textrm{HPWL} \\
&& \textrm{s.t.}\,\,\, \mathbf{T}  = \mathbf{A}^{-1}\mathbf{b}\nonumber \\
&& \textrm{s.t.}\,\,\, \textrm{p-norm}(\mathbf{T})-T_{th} \le 0 \nonumber \\
&& \textrm{s.t.}\,\,\, g_0(\mathbf{p}) \le \nonumber 0.
\end{eqnarray}
Remarkably, this optimization converges in fewer than 15 iterations. The final configuration, depicted in Fig.~\ref{fig:map}c, maintains the symmetry of the first guess, analogously to Case 2. The optimization finds a balance between two contrasting needs: Minimizing the maximum temperature requires chiplets to be spaced apart, while minimizing the HPWL necessitates closer packing of the dies. This trend is illustrated in Fig.~\ref{fig:case_3}a and b, where the HPWL increases and the maximum regularized temperature decreases until it falls below $T_{th}$, stabilizing around 76.28°C $<T_{th}$. The corresponding real maximum temperature is 80.81°C, which is below the intended target. The \texttt{ANSYS}'s prediction is within 1\% error (81.67°C). The actual HPWL for the final configuration is 128.6 mm, which, as expected, lies between Cases 1 and 2. As depicted in Fig.~\ref{fig:case_3}c, the nonoverlap constraint is initially violated due to the need for a low HPWL but eventually falls below zero. This behavior differs from Case 2, where the nonoverlap constraint does not conflict with other requirements. We note that in practical layouts chiplets are packed tighter in order to satisfy electrical constraints, such as data delay~\cite{chen2023floorplet}. Possible future iterations of \texttt{DiffChip} will include these constraints. Additionally, the orientation of the chiplet can be considered a degree of freedom. Since changing orientation, such as rotating by 90°, is nondifferentiable, regularization will be necessary.

\section{Conclusions}\label{conclusions}
In conclusion, we present \texttt{DiffChip}, a framework for chiplet floorplanning based on automatic differentiation (AD). By integrating a differentiable thermal solver with smooth space-dependent properties such as thermal conductivity and heat sources, we compute the gradients of the cost function with respect to the chiplet positions. Additionally, the maximum temperature and half-perimeter wirelength (HPWL) are made differentiable through regularization. We successfully apply \texttt{DiffChip} to various scenarios, including minimizing the HPWL under maximum temperature and non-overlap constraints. The final layouts are validated against a commercial software. Unlike gradient-free methods, such as simulated annealing, our approach achieves faster convergence and enables large-scale optimization. Lastly, unlike machine learning approaches, \texttt{DiffChip} requires neither training nor a surrogate, as the underlying physical model is inherently differentiable.

\section*{Acknowledgment}

The authors thank Arvind Kumar from IBM for helpful discussions and support.

\bibliographystyle{IEEEtran}
\bibliography{IEEEabrv,biblio}

\end{document}